\documentclass[twocolumn,showpacs,amsmath,amssymb,superscriptaddress,floatfix]{revtex4-1}
\pdfoutput=1

\usepackage{amsmath}
\usepackage{amsfonts}
\usepackage{float}
\usepackage{ graphicx } 
\usepackage{ bm } 
\usepackage{ color }
\usepackage{epstopdf}
\usepackage{caption}
\usepackage{subcaption}
\newcommand{\bea}{\begin{eqnarray}}
\newcommand{\eea}{\end{eqnarray}}

\begin{document}

\title{Novel Approach to Unveil Quantum Phase Transitions Using Fidelity Map}
\author{Ho-Kin Tang}
\email{hokin@csar.ac.cn}
\affiliation{Shenzhen JL Computational Science and Applied Research Institute, Shenzhen, China}

\author{Mohamad Ali Marashli}
\affiliation{Department of Physics, City University of Hong Kong, Kowloon, Hong Kong}

\author{ Wing Chi Yu }
\email{wingcyu@cityu.edu.hk}
\affiliation{Department of Physics, City University of Hong Kong, Kowloon, Hong Kong}

\date{\today}

\begin{abstract}
Fidelity approach has been widely used to detect various types of quantum phase transitions, including some that are beyond the Landau symmetry breaking theory, in condensed matter models. However, challenges remain in locating the transition points with precision in several models with unconventional phases such as the quantum spin liquid phase in spin-1 Kitaev-Heisenberg model. In this work, we propose a novel approach, which we named the fidelity map, to detect quantum phase transitions with higher accuracy and sensitivity as compared to the conventional fidelity measures. Our scheme extends the fidelity concept from a single dimension quantity to a multi-dimensional quantity, and use a meta-heuristic algorithm to search for the critical points that globally maximized the fidelity within each phase. We test the scheme in three interacting condensed matter models, namely the spin-1 Kitaev Heisenberg model which consists of the quantum spin liquid phase and the topological Haldane phase, the spin-1/2 XXZ model which possesses a Berezinskii–Kosterlitz–Thouless transition, and the Su-Schrieffer–Heeger model that exhibits a topological quantum phase transition. The result shows that the fidelity map can capture a wide range of phase transitions accurately, thus providing a new tool to study phase transitions in unseen models without prior knowledge of the system's symmetry.
\end{abstract}

\maketitle

\section{Introduction}
The study of QPTs has been one of the hottest research topics in modern condensed matter physics. At absolute zero temperature, the ground state wavefunction of a many-body system can undergoes an abrupt change in its qualitative structure when the external driving parameter varies. Such a transformation is completely driven by the quantum fluctuations in the ground state, and is known as the quantum phase transitions (QPTs) \cite{Sachdev2000,Carr2010}. The quantum critically arise from a QPT can give rise to many interesting collective phenomena even at finite temperatures. Understanding QPTs is believed to play a key role in understanding the mechanism of unconventional superconductivity (e.g. \cite{Wu2011,Mathur1998}). Moreover, due to the advance in quantum technologies, theoretically proposed condensed matter models can now be realized and studied in a controlled manner using platforms such as cold atoms, optical lattices, and superconducting qubits \cite{Georgescu2014}, making experimental verifications of theoretical proposals possible.

Traditional approaches to study QPTs is via the Landau's spontaneous symmetry breaking theory in which the local order parameter plays the key role. However, finding the suitable order parameter that characterizes the phase transition is a highly non-trivial task. Especially for transitions such as the topological phase transitions and the Berezinskii-Kosterlitz-Thouless (BKT) transitions in which the order parameter is usually non-local or even does not exist. On the other hand, concepts from quantum information science such as the fidelity \cite{Quan2006,Zanardi2006,Gu2010}, quantum entanglement \cite{Osterloh2002,Osborne2002,Amico2008}, and quantum discords \cite{Dillenschneider2008,Sarandy2009,Werlang2010} have been borrowed to study QPTs in the recent two decades and find a great triumph in detecting the phase boundaries in many condensed matter models . The application of these schemes do not require any a prior knowledge in the system's symmetry, thus making them an advantageous indicator of the phase transitions over the conventional method.

\begin{figure*}[htb!]
\centering

\includegraphics[width=0.95\textwidth]{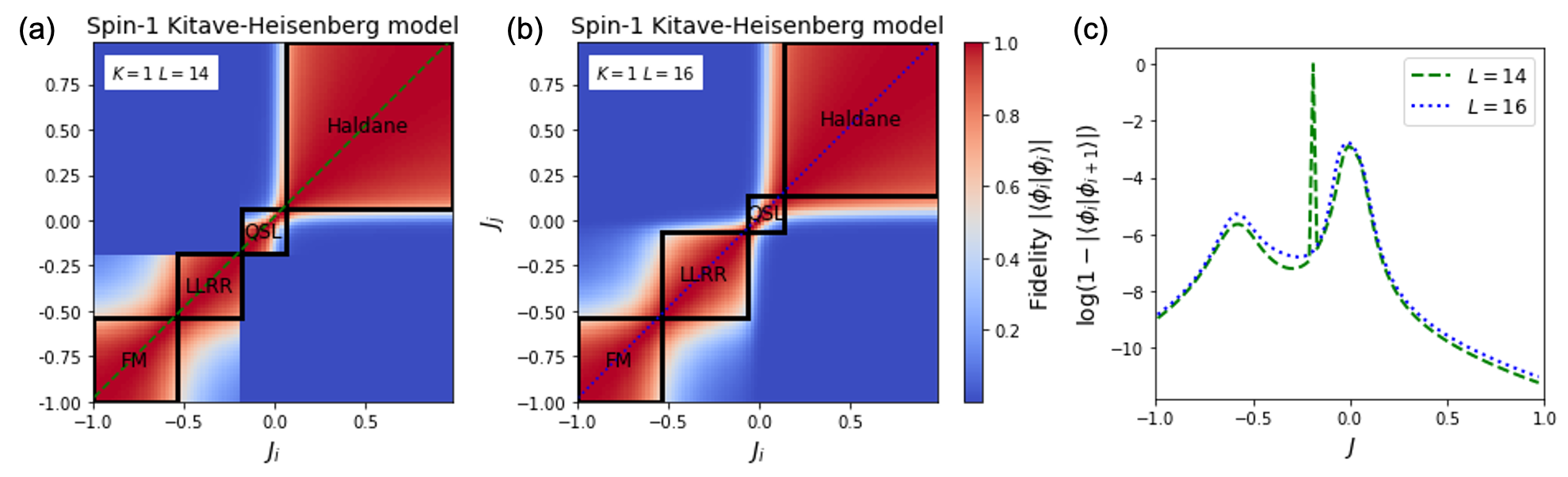}
\caption{Detecting the quantum spin liquid (QSL) with the fidelity map in the spin-1 Kitaev Heisenberg model. With varying parameter $J$, we construct the fidelity map of the model for (a) $L=4n+2$ sites and (b)$L=4n$ sites. The red color indicates the high fidelity region while the blue color indicates the low fidelity region. The squares with black outline are the optimized solution we obtained from maximizing the fidelity sum using the cuckoo search. (c) The conventional measurement of logarithmic of 1-fidelity, that is measured along a dashed line in (a) and (b). (FM: ferromagnetic phase, LLRR: left-left-right-right phase)}
\label{fig:KH}
\end{figure*}

The fidelity is defined by the overlap between the ground state wavefunctions that are differed by a small value in the driving parameter. As the ground state wavefunction on two sides of a QPT is qualitatively different, one shall expect the fidelity shows a sudden drop across the transition while attains a value close to one within the same phase. The fidelity, and its variants such as the fidelity susceptibility, has been proven to be capable in detecting the quantum phase transition in a number of condensed matter models such as the one-dimensional (1D) and two-dimensional (2D) transverse-field Ising models, 2D XXZ model \cite{Yu2009}, the Lipkin–Meshkov–Glick model \cite{Kwok2008}, XY model \cite{Zanardi2006}, 2D Kitaev honeycomb lattice model \cite{Yang2008,Wang2010}, just to name a few. An equality between the fidelity susceptibility and the spectral weight has also been established suggesting that the fidelity susceptibility can be experimentally measured using neutron scattering or the ARPES techniques \cite{Gu2014}. However, whether the fidelity approach can be used to detect a BKT transition is still a controversial problem. For examples, it was found that the fidelity fail to detect the BKT transition in the 1D fermionic Hubbard model \cite{You2007,Gu2008}. Similarly in the 1D Bose-Hubbard model, the fidelity susceptibility's peak was found to deviate from the exact BKT transition point of the model \cite{Carrasquilla2013,Lacki2014}.

In this work, we introduce a fundamental extension to the fidelity to detect QPTs. We use the overlap between different ground states among a parameter range to construct a two dimensional map that we name as the fidelity map. Comparing to conventional fidelity approaches~\cite{Gu2010}, our scheme shares the same advantage of detecting QPTs without prior knowledge of system, but utilize more information from the system. We are not only considering the fidelity between ground states with small changes in the driving parameter, but among the whole parameter space. This allows us to gain more information about the system. Moreover, the similarity between the ground state in different phases is visually clear to recognize in the map. If the ground state across the QPT is completely different, a very sharp distinguishable region will be shown on the fidelity map. However, if two concerned phases shares some similar features, the transition would occur rather gradually on the fidelity map.

The fidelity map contains a great amount of information for searching QPTs. We can often find recognizable regions on the map that represent the quantum phases. But when gradual changes of fidelity map between regions happens, it requires some objective function to find where QPT occur. We choose the method of using square boxes on the map and optimizing the fidelity within the boxes. First we use a square box to define each individual phase, in which we expect the fidelity between the ground states is high consistently. Outside the box, the fidelity between ground states is low as they belong to different quantum phases. Our ultimate target is to maximize the fidelity sum in all boxes. We define our objective function as the fidelity sum in all boxes and the variables and the QPT points. We use a meta-heuristic algorithm, cuckoo search, to perform the task, carrying two benefits. First, the technique is able to find global minimum without trapping in local minimum. Second, the technique is much more efficient in optimization comparing to the brute-force approach, especially when the number of variables increases, i.e. more QPTs occurs in the model. We would give details of the method in the next section.

In this manuscript, we use the novel approach of fidelity map to explore the QPTs in three representative models. The approach combines both the concept of fidelity map and the optimization technique using a meta-heuristic algorithm. We first find the ground states using the exact diagonalization technique, then we construct the fidelity map accordingly. In the spin-1 Kitaev Heisenberg model, we identify the location of three QPTs points. We also find a region on the fidelity map that the controversial quantum spin liquid phase occurs, in which no order parameter is available to detect it at presence. In spin-half XXZ model, we find the occurrence of Berezinskii–Kosterlitz–Thouless transition between XY phase and Haldane phase. In spinless SSH model, we demonstrate the ability of fidelity map to identify topological phase transitions. The manuscript is organized as the following. In section II, we discuss the approach of fidelity map in details. In section III, we show its application in three representative models, and demonstrate its ability in detecting different types of QPTs. In section IV, we give a conclusion.

\section{methods}
\subsection{The Fidelity map}
Given a quantum many-body system described by the Hamiltonian $H(\lambda)$, where $\lambda$ is the driving parameter, we define the fidelity map as
\begin{equation}
    F(\lambda_i,\lambda_j) =  \vert \langle \phi(\lambda_i) \vert \phi(\lambda_j)\rangle \vert
\end{equation}
, where $\vert \phi(\lambda_i)\rangle$  is the ground state of the system, with $H(\lambda)\vert \phi(\lambda)\rangle = E_0(\lambda) \vert \phi (\lambda) \rangle$. $F(\lambda_i,\lambda_j)$ measures the similarity of the ground states in the system with different driving parameter $\lambda_i$ and $\lambda_j$. The conventional  fidelity approach that considers only the overlap between ground states of neighbor parameter, $\vert \langle \phi (\lambda) \vert \phi (\lambda+\delta \lambda) \rangle \vert$ for small $\delta \lambda$. Our approach of fidelity map is different from the conventional fidelity. $\vert \langle \phi(\lambda_i) \vert \phi(\lambda_j)\rangle \vert$ considers any two different ground states in the parameter space, in which $\vert \lambda_i - \lambda_j \vert$ can be much larger than $\delta \lambda$ in the fidelity approach. The fidelity considers the similarity between all ground states in the parameter space. For simplicity, we notate $\vert \phi(\lambda_i)\rangle$ using $\vert \phi_i\rangle$ in the following text.

Take Fig.~\ref{fig:KH}(a) as an example, it shows the fidelity map of the spin-1 Kitaev Heisenberg model. The fidelity map covers whole parameter range, while the conventional fidelity approaches only consider the fidelity along dashed line. Thus, the fidelity map provides us a more complete picture to pin down the phase transition point accurately. Within the same phase, the ground states would have higher fidelity with each other, thus forming a regime that is visually recognizable.

\subsection{Objective function}
In the fidelity map, some of the phases are easily identified as a square box visually, but some of them are not readily distinguishable due to the similarity with their neighboring phase, like the FM to LLRR phase transition in Fig.~\ref{fig:KH}(a). We only observe the gradual narrowing of the high fidelity region between two phases. To identify the QPTs, we use the square bounding box for each phase, and we sum the fidelity within the boxes. 

\begin{equation}
    C(\{\bm{k}\})=\sum_{\{\bm{k}\}} \sum_{i,j \in \bm{k}} F(\lambda_i,\lambda_j)  
\end{equation}
, where $\{\bm{k}\}$ defines the location of the boxes. The cost function is defined as the total fidelity sum within all boxes. The target of the optimization is to maximize the fidelity within all the boxes. Here, we assume that the fidelity within the same phase is maximized when we choose the box that the parameter range overlaps with the quantum phase.

\subsection{Cuckoo Search}
The optimization technique is required to search for the boxes, as the number of combination locating the phase transition increases exponentially with the number of phase transition and the precision of our measurement. To efficiently perform the task, we use a population-based meta-heuristic algorithm called cuckoo search~\cite{Yang2009}, which is inspired by the breeding behavior of cuckoos. It is one of the famous evolutionary algorithms~\cite{Simon2013}. In the cuckoo search, the searching process is by a Levy flight that mimics the flying pattern of birds. The Levy flight balances between efforts in the exploration and the exploitation in searching process, thus the algorithm gains efficiency in finding the optimized solution without being trapped in local minimums. In the algorithm, a cuckoo in the population will replace others eggs if it used the Levy flight to find a better solution, therefore achieving better and better quality of solution after iterations, eventually reaching the global minimum. 

In our analysis scheme, we use cuckoo search to find optimized solutions in a discrete space of QTP points. The cuckoo search serves two important purposes in optimization. The first one is that it can escape local minimum in the cost function. Other methods like the gradient method are easily getting trapped in the searching process, while the cuckoo search can avoid it using the heavy-tail Levy flight, allowing global searching even already approached local minimum. The second one is the efficiency in searching the solution. The function space exponentially increases with the number of QTPs and the precision of our measurement. Searching optimized solution with brute force also requires exponential amount of time. Cuckoo search provides a way to find a optimized solution efficiently.  

\section{Applications}
\subsection{Non-trivial quantum spin liquid: Spin-1 Kitaev Heisenberg model}

The Hamiltonian of the 1D Kitaev Heisenberg model is given by
\begin{eqnarray}
    H&=& K\sum_j^{L/2} (S_{2j-1}^x S_{2j}^x + S_{2j}^y S_{2j+1}^y)\\ \nonumber
    &&+ J \sum_j^L \bm{S_j \cdot S_{j+1}} -h\sum_{j=1}^LS_j^z,
\end{eqnarray}
where $\mathbf{S}_j = \{S_j^x,S_j^y,S_j^z\}$ are the spin-1 operators at site $j$ satisfying the SU(2) alegbra. The parameter $K$ and $J$ characterises the Kitaev coupling and the Heisenberg coupling strength respectively and $h$ is the external field strength. For $K=0$, the model reduces to the well-known spin-1 Heisenberg model and previous study shows a QPT from the topological Haldane phase to a topologically trivial magnetic phase takes place at $h\sim 0.42$ when the field increases from below \cite{Haldane1983}. For $K\ne 0$ and $h=0$, the ground state phase diagram is found to consists of four phases, namely the ferromagnetic phase, the LLRR phase with spin configuration of the form $|\rightarrow\rightarrow\leftarrow\leftarrow\cdots\rangle$, a gapped quantum spin liquid phase, and a topological Haldane phase \cite{You2020}. In the following, we will focus on the later case.

Figure \ref{fig:KH}(a) and (b) shows the fidelity map of the Kitaev Heisenberg model for $L=14$ and $L=16$ respectively. In both cases, four high fidelity regimes as indicated by the red color can be resolved from the map. With the cuckoo search method introduced in the previous section, the optimal solution gives the transition points located at the top right concern of the black boxes. The obtained transition points agree quite well with the theoretically values, i.e. $J_{FM-LLRR}=-0.6, J_{LLRR-QSL}=-0.08$ and $J_{QSL-Haldane}=0.08$, proposed in Ref.~\cite{You2020}.

We compared the fidelity map approach with the conventional fidelity measure in detecting the QPT in the model and find that the fidelity map actually out-performs the conventional fidelity approach. Figure \ref{fig:KH}(c) shows the conventional fidelity measure using $|\langle\phi_0(J)|\phi_0(J+\delta J)\rangle|$ as a function of $J$. For $L=16$, only two peaks are observed. The peak around $J=-0.6$ detects the transition between the FM and the LLRR phases. However, there is only one broad peak occurring around $J=0$ and the transition to and from the QSL phase can not be resolved. On the other hand, for $L=14$, a sharp peak appears in the close vicinity on the left of the broad peak around $J=0$. While this may indicates a transition between the LLRR and the QSL phase, we can see that the location of this sharp peak and the broad peak's maximum is not symmetric about $J=0$, which is different from what was theoretically suggested.

\subsection{BKT Transition: 1D spin-1/2 XXZ model}
Berezinskii–Kosterlitz–Thouless~(BKT) phase transition is a quantum phase transition of infinite order~\cite{Kosterlitz1989}. Near the transition point, the correlation function decays algebraically, instead of decaying exponentially in the second order phase transition. It is controversial whether the conventional fidelity susceptibility can probe the BKT transition, as the fidelity susceptibility captures only the leading few terms in fidelity. We test our scheme in 1D spin-1/2 XXZ model~\cite{Kargarian2008, Yao2012}, where the Hamiltonian reads as
\begin{equation}
    H= \sum_i (S_i^x S_{i+1}^x + S_i^y S_{i+1}^y) +\Delta \sum_i S_i^z S_{i+1}^z,
\end{equation}
where $\Delta$ determines the anisotropy in the model, with $\Delta=1$ as Heisenberg model limit. At the limiting case of  $\Delta\rightarrow \infty$ or $\Delta\rightarrow -\infty$ , either the ferromagnetic order or Neel order is the dominant phase, as the third Ising term dominates. Between two phases, the first two exchange terms extend the quantum fluctuation in the system, resulting in the intermediate XY term. BKT transition is known to occur at $\Delta=1$. 

\begin{figure}[h]
\centering
\includegraphics[width=0.4\textwidth]{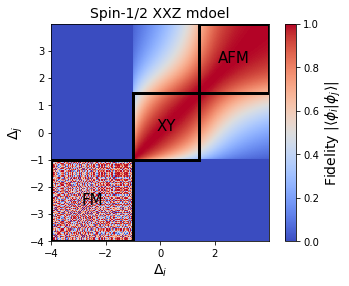}
\caption{Berezinskii–Kosterlitz–Thouless transition in the spin-1/2 XXZ model. We construct the fidelity map by varying the anisotropic parameter $\Delta$. The red color indicates the high fidelity region while the blue color indicates the low fidelity region. We find that the high fidelity region becomes narrower between the XY phase and the anti-ferromagnetic AFM phase, in which our analysis scheme gives the quantum phase transition point with black boxes. In ferromagnetic FM phase, the noise feature in fidelity map indicates the existence of degenerate ground states. The lattice size of the model is $L=18$ .}
\label{fig:XXZ}
\end{figure}

Figure \ref{fig:XXZ} shows the fidelity map of the spin-1/2 XXZ model. Using the cuckoo search, three distinct phases are found and the parameter regime in which each of the phases exists is framed by the squares with black outlines. The ferromagnetic to XY phase transition occurs exactly at $\Delta=-1$, which matches the theory prediction. We observe the narrowing of the high fidelity regime occurs at the XY to anti-ferromagnetic transition. Our analysis scheme gives the phase transition point $\Delta_c \approx 1.5$. The derivation from the exact result $\Delta_c=1$ might be due to the small system size we are using, thus could be further improved when we have the result of larger lattice size. In addition, the noise feature of the fidelity map can also help identify the ferromagentic phase as a degenerate ground state.

\subsection{Topological phase: spinless SSH model}

\begin{figure}[h]
\centering
\includegraphics[width=0.45\textwidth]{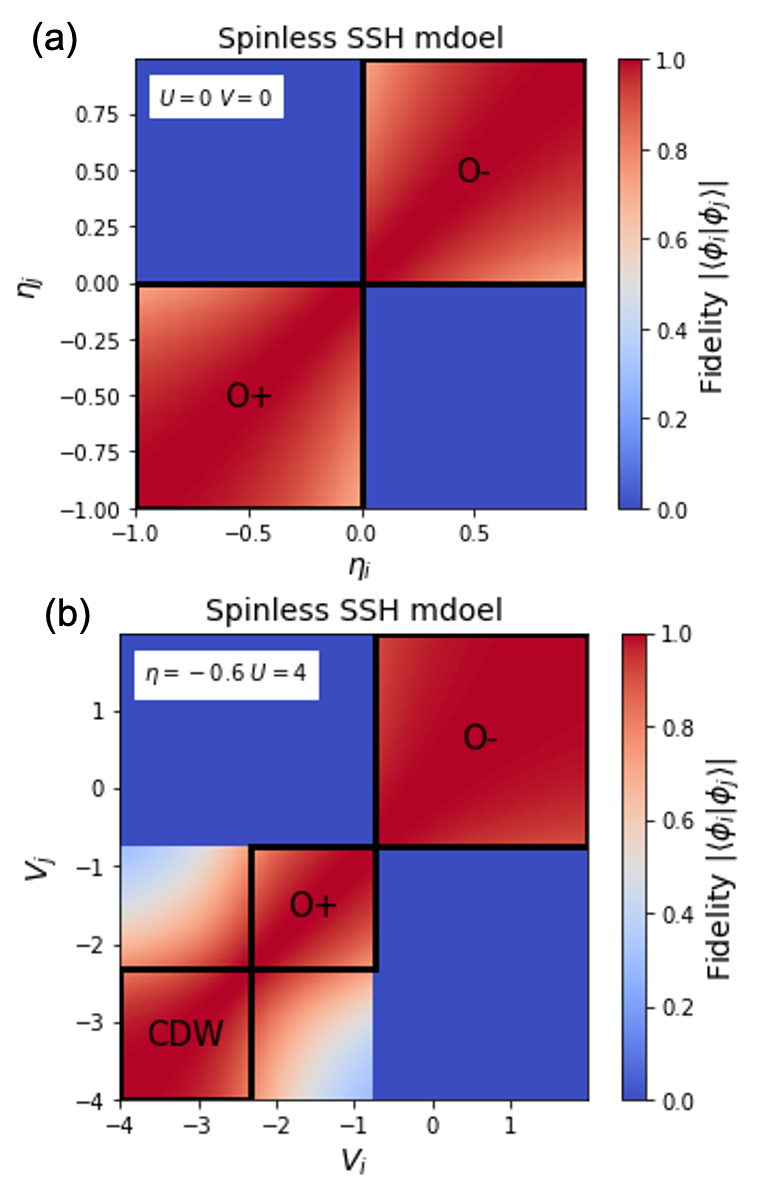}
\caption{Detecting topological phase transitions in the spinless Su-Schrieffer–Heeger model. We shows the fidelity map of (a) the non-interacting model varying anisotropy parameter $\eta$, and (b) the interacting model varying nearest neighbor interaction $V$ with fixed $U=4$ and $\eta=0.6$. The fidelity map together with our analysis scheme succeeds to capture the topological phase transition from the topological trivial phase~(O+) to the topological non-trivial phase~(O-). In the fidelity map, the red color indicates the high fidelity region while the blue color indicates the low fidelity region. The lattice size of the model is $L=12$. (CDW: charge density wave.)}
\label{fig:SSH}
\end{figure}

The Hamiltonian of the Su-Schrieffer–Heeger model with interactions reads
\begin{eqnarray}
H& =& -t\sum_j [(1+\eta)c^\dagger_{j,A} c_{j,B} + (1-\eta)c^\dagger_{j,B} c_{j+1,A} +h.c.] \\ \nonumber
&&+U\sum_{j} n_{j,A}n_{j,B} + V\sum_{j} n_{j,B} n_{j+1,A},
\end{eqnarray}
where $t$ is the hopping amplitude, $\eta$ is the parameter charactersing the anisotropy in intercell and intracell hopping, $U$ and $V$ characterizes the strength of intracell and intercell interactions respectively. Unless otherwise specified, we take $t=1$ and adopt periodic boundary condition in our simulation. In the absence of interactions, the Hamiltonian can be exactly diagonalized in the Majorana fermion picture. The system is in a topological phase for $\eta<0$ in which there are decoupled Majorana fermions at the two edges, while it is in a trivial phase when $\eta>0$ and the all the Majorana fermions are paired up. A topological quantum phase transition takes place at $\eta=0$ \cite{Su1979,Yu2016}.

In the presence of interactions, the model exhibits a rich ground state phase diagram. By analysing the entanglement entropy and the order parameter derived using the mutual information method proposed by Gu et. al.~\cite{Gu2013}, the topological phase was found to be robust against the addition of repulsive intercell interactions and exists in the positive $\eta$ regime when appropriate interactions are introduced. Besides the topological phase and the trivial phase, it is also found that the model posses two types of charge density wave (CDW) phases and phase separation phase in different parameter regimes \cite{Yu2016}.

Figure $\ref{fig:SSH}$ shows the fidelity map in the SSH model with and without interactions for a system size of $L=12$. In Fig.~\ref{fig:SSH}(a), two squares in red can be easily recognized by bare eyes. The ground state with the parameter values within each of the squares are highly similar to each other while the fidelity dropped to almost zero across the squares. This clearly indicates that there are two distinct phases. Using cuckoo search, the transition point is found to be located at $\eta=0$, agreeing with well-known result.

Figure $\ref{fig:SSH}$(b) shows the case with interactions. Using cuckoo search, three distinct phases are found and the parameter regime in which each of the phases exists is framed by the squares with black outlines. The transition points located by the vertexes of the squares along the diagonal agree roughly with the one obtained in Ref.~\cite{Yu2016}. Small derivation exists due to the finite size effect. From the fidelity map, we can also tell the "likeness" between two phases. When going between the topological phase and the trivial phase, the two corresponding red squares of high fidelity regime are clearly distinguishable. However, when going between the trivial phase and the CDW phase, the fidelity decreases more gradually outside the framed square. This suggest that the CDW phase and the trivial phase are more alike as compared to the trivial and the topological phase.

\section{Conclusion}

In conclusion, we have fundamentally extended the concept of fidelity to the novel fidelity map, in which we consider the fidelity between ground states among a parameter range instead of just neighboring parameter. We analyze the fidelity map by maximizing the sum of fidelity within the bounded box in individual phase. Thus we identify where quantum phase transition happen without prior knowledge of the system. The proposed scheme works well in three representative systems. Our scheme identifies the transitions to and from the quantum spin liquid phase in the spin-1 Kitaev Heisenberg model, that the conventional fidelity approaches fails. Also, our scheme works in both BKT transitions and topological phase transitions. Moreover, the fidelity map is highly visual recognizable, thus showing the similarity between the ground states of different phases. For further research directions, we can enhance the performance of the fidelity map using different combination of objective functions and bounding methods, and verify our approach in the open system~\cite{Breuer2002}, in which the fidelity map between the time evolved ground states may help identify the phase transition.

\section{Acknowledgement}
H.K.T. thanks Noah Yuan for the helpful discussion. We acknowledge financial support from Research Grants Council of Hong Kong (Grant No. ECS/21304020) and City University of Hong Kong (Grant No. 9610438).

\end{document}